# Assembly and disorder dissipation in superparamagnetic nanoparticle chains in a rotating magnetic field

Zhixing He[a] and Hans D. Robinson[a*]

[a]Virginia Tech, Department of Physics, Blacksburg, VA, USA 24061

* Corresponding author; hansr@vt.edu

We investigate the formation of chains of superparamagnetic iron oxide nanoparticles (SPIONs) in a rotating magnetic field, combining two well-explored chain-forming systems: larger micron-scale beads in a rotating magnetic field, and SPIONs in a static field. This simple combination is interesting because it features self-assembly that occurs both far from equilibrium and at a finite temperature, with the better-explored systems constituting respectively its zero temperature and near-equilibrium limits. Theories applicable to either of the two limits qualitatively predict the chain length distributions, except that chains in our experiments are shorter, which we attribute to the simultaneous presence of thermal fluctuations and fluid shear forces that work in concert to break chains apart. Our most striking result is that the disorder in the SPION chains gradually dissipates over a timescale of tens of minutes, about two orders of magnitude slower than the characteristic chain assembly time. The disorder dissipation can be sped up by increasing particle concentration and solution ionic strength, both of which increase the speed of chain assembly. This strongly suggests that the improvement in chain order with time is not due to thermal fluctuations but rather to energy imparted by the self-assembly process, which continually causes chains to grow and break apart, even when a steady state distribution has obtained. More generally, our results indicate that self-assembly away from equilibrium may sometimes lead to better ordered assemblies than under near-equilibrium conditions.

## I. INTRODUCTION

Superparamagnetic iron oxide nanoparticles (SPIONs) are a class of magnetic nanoparticles that exhibit strong directional magnetic interactions that can be conveniently turned on and off with an external field. This makes them particularly interesting for fundamental studies of nanoassembly, where the linear chains they form under the application of a magnetic field can serve as a model and proving ground for more complex particle assemblies, and also are useful as coarse-grained models for molecular polymers studied [1, 2]. SPIONs and similar superparamagnetic particles have also been investigated for potential applications in a variety of fields, such as drug delivery [3, 4], MRI contrast agents [5, 6], cancer theranostics [7-9], nanoscale mixing applications [10, 11], nanophotonics [12, 13], and nanocatalysis [14].

The assembly of SPIONS into linear chains [15, 16] as well as more complex structures [15, 17] in a static magnetic field have been extensively studied, and a mature body of theory with good predictive power exists for predicting the length and size distributions in such assemblies [17]. The development of this theory has been facilitated by the fact that in a static field, equilibrium statistical mechanics methods can be used to arrive at expressions of wide applicability. On the other hand, if the magnetic field were to be varied rapidly in time, the foundations of this theory is removed, as the assembly does not take place near equilibrium, and it is no longer possible to assign a well-defined potential energy to a given configuration of particles.

If we turn instead to larger superparamagnetic beads, whose diameters typically are in the μm range, there is a large and growing body of work on particle assembly in time-varying fields, particularly in fields that rotate with a constant angular velocity $\omega$ [18-20]. Because this assembly takes place away from thermal equilibrium, these systems provide many examples of rich and unexpected phenomena. For example, when chains grow beyond a certain length, they tend to fracture near their center [20], resulting in two daughter chains of half the length of the original chains. If the particles in the chains are permanently bound, they instead bend into folded assemblies when their length exceed a critical value [21]. Oscillating magnetic fields can also be used to assemble and drive magnetic swimmers made up of magnetic beads [22-24]. Theoretical progress in this area has also been significant, in large part because the microbeads are large enough that the effects of temperature can be ignored, and the modelling of observed phenomena can assume completely deterministic behavior. Understanding of these systems have also been aided by their size, which makes it possible to directly observe particle assembly with optical microscopy, and by the very low disorder that is often seen in these systems.

In this study, we will combine both of these regimes, and study the assembly of SPIONs in a rotating magnetic field. Assembly in this system combines the finite temperature and far-from equilibrium features of the better-studied regimes, which can therefore be seen as respectively its deterministic (or zero temperature) and near-equilibrium limits. The simplicity of the system makes it a good candidate for studying phenomena far from equilibrium and non-zero temperatures, but we face the issue that detailed *in situ* monitoring of the assembly is generally not possible, neither with optical nor other methods, except that ensemble averages of some quantities can sometimes be obtained.

As we will see, we are able to make progress in spite of these difficulties. Notably, we overcome the imaging problem by making permanent any chains that form in the magnetic field through encapsulation in silica using a modified Stöber process [25], so that the chains can be imaged with transmission electron microscopy (TEM) outside of the magnetic field space. The encapsulation occurs sufficiently rapidly that the effect of changing the incubation time in the field prior to encapsulation by as little as five minutes has a detectable effect on chain morphology.

Performing the assembly under a variety of conditions, we first note that the SPION chains produced in our experiments share features with both the better explored regimes. In particular, the length distribution takes the form of a decaying exponential, just as is the case for SPION chains assembled in a static field. Similarly, the maximum chain length varies as $1/\omega$, just as it does for larger magnetic beads assembled in a rotating magnetic field. However, there are also important differences. In particular, the chains in our experiments are shorter than predicted by the theory applicable to SPIONs in a static field and by the theory for micro-bead assembly in a rotating field.

Next, we turn to examining the disorder in the chains, which is an important topic in self-assembly in general [26, 27], and a known issue in SPION chains in particular [28]. When structures self-assemble from a disordered collection of smaller constituents, it is often crucial for the intended applications that disorder is kept to a minimum. Unfortunately, this does not always occur. Even though there may be a specific configuration that corresponds to a clear global minimum in free energy—for superparamagnetic particles in a magnetic field this minimum consists of a straight colinear chain of particles in direct contact—the constituents are very likely to initially assemble in ways that differ, subtly or significantly, from the ideal structure. This non-ideal assembly may relax toward the global minimum, but may also get stuck in a local minimum, which, if it is deep enough, could be permanent. This phenomenon is known as kinetic trapping [29], and is one of the greatest impediments to the efficient use of self-assembly in bottom-up fabrication applications.

In our case, we find that we are able to substantially ameliorate the disorder present in the SPION chains when they first form. We show that this can be accomplished most simply by allowing the particle dispersions to incubate in the rotating magnetic field for up to two hours before encapsulation. We also find that this disorder healing proceeds faster if salt is added to the dispersion, or if the concentration of particles is increased. The chain disorder in all three cases is improved in a very similar manner, even while the process modifications that produce the improvements have little or no appreciable effect on the chain length distribution. We attribute the disorder dissipation to the dynamic steady state, where all chains are continually growing by merging with particles and other chains, and also regularly undergo scission spontaneously or through collisions with other chains. We believe these processes provide energies that are large compared to the thermal energy, allowing disorder to dissipate as the system moves toward a more ordered equilibrium state, even when the disordered local minimum is too deep for thermal fluctuations to dislodge it.

## II. Results and discussion

Before discussing our results, we will give a brief background to the two limiting cases of assembly of SPIONs in a static magnetic field and of magnetic microbeads in a rotating field.

### A. Background

#### 1. SPION assembly in a static magnetic field.

SPIONs are single crystalline nanoparticles made of ferromagnetic iron oxide compounds such as magnetite ($Fe_3O_4$). Magnetically, they are characterized by an easy axis, along which the particle magnetization is preferentially oriented, and a perpendicular hard plane that presents an energy barrier $E_{an} = KV$ against magnetization reversal. Here, $V$ is the volume of the nanoparticle, and $K$ is the magnetic anisotropy energy density of the material. If $E_{an}$ is small enough, thermal fluctuations are able to randomly flip the particle magnetization, which occurs at a characteristic time scale known as the Néel relaxation time [30, 31]:

$$\tau_N = \tau_0 \exp\left(\frac{KV}{k_B T}\right). \tag{1}$$

The attempt time $\tau_0$ is typically in the range of $10^{-9}$ or $10^{-10}$ s. If $\tau_N$ is small on timescales that are relevant for diffusion of the particles (typically a few µs for SPIONs), magnetization fluctuations will be rapid enough that well-separated particles will exert no net magnetic force on each other, even though the particles are ferromagnetic and carry a permanent magnetic moment. This leads to the phenomenon known as superparamagnetism. From Eq. (1) we see that $\tau_N$ is very sensitive to the particle volume, so superparamagnetism only occurs in particles that are smaller than a certain size, which for iron oxide particles in water is approximately 25 nm in diameter [32]. The particles we use here are about 21.2 nm in diameter, and therefore in the superparamagnetic regime.

When an external magnetic field $B$ is applied to a SPION, it will preferentially align its easy axis with the magnetic field, spending most of the time in the magnetic state that is aligned with the direction of the applied field. Averaged over a time $t \gg \tau_N$, the particle therefore acquires a net magnetization, which in the limit of $\frac{KV}{k_B T} \gg 1$ is given by [33]

$$M = M_S \tanh \frac{M_S V B}{k_B T}, \qquad (2)$$

where $M_S$ is the saturation magnetization per unit volume for the SPION material. In the limit of $\frac{KV}{k_B T} = 0$, the hyperbolic tangent is replaced by the Langevin function $L(\xi) = \coth \xi - 1/\xi$, but we will use Eq. (2) since it is a somewhat better approximation for the particles we work with here.

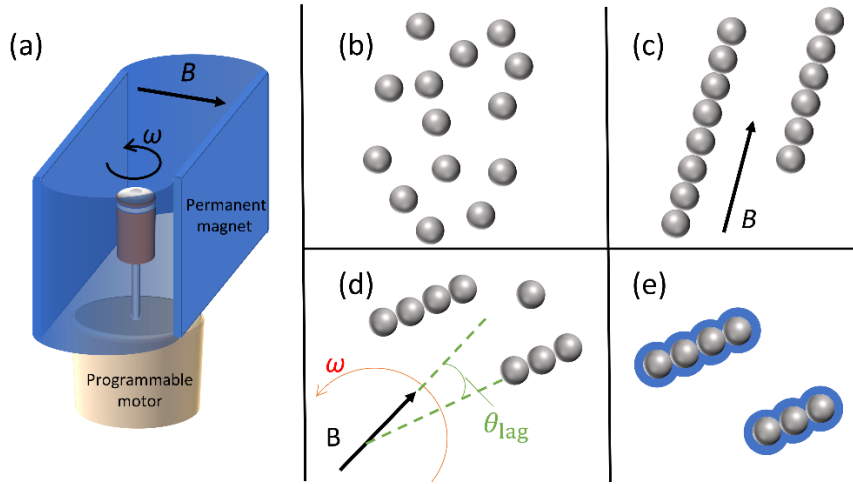

Figure 1: (a) Schematic diagram of the experiment. The sample vial is mounted to the drive of the programmable motor, placing it at the center of a uniform magnetic field generated by two permanent magnets. (b)-(e) Outline of the fabrication of the nanocolloidal chains: (b) In the absence of a magnetic field, the citric acid-coated SPIONs have a purely repulsive interaction and remain well dispersed; (c) In a static uniform magnetic field, SPIONs align into chains; (d) In a rotating field, chain length is capped due to hydrodynamic drag on the rotating chains; (e) chains are encapsulated in silica coating to preserve them after the field is switched off.

The interaction energy between magnetic dipoles that are oriented in the same direction, separated by a distance $r$, and aligned at an angle $\theta$ is

$$E_B = -\frac{\mu_0 (MV)^2}{4\pi r^3}(3\cos^2\theta - 1) \qquad (3)$$

where $M$ is given as magnetic moment per unit volume. The equation is equally valid for spherical particles, where $r$ is the center-to-center distance. In other words, the interaction is attractive when $\theta = 0$ (when dipoles are aligned end-to-end), but repulsive when $\theta = \frac{\pi}{2}$ (side-to-side alignment). Once $B$ is large enough that $E_M$ becomes comparable to $k_B T$, the particles will therefore begin to assemble into a linear chain where the chain as a whole as well as the easy magnetic axes of each constituent particle is aligned with the external field (Fig. 1). If the $E_B \gg k_B T$, the chain will in principle grow indefinitely, limited in length only by the number of particles available and the size of the containing vessel. For practically obtainable SPIONs, this condition is not met, and even when $M = M_S$, $E_B$ is only tens of $k_B T$. As a result, any chain can be broken by a thermal fluctuation, and the frequency of fragmentation increases the longer the chain gets. This process takes place close to thermal equilibrium, so standard techniques from statistical mechanics can be applied to show that at steady state, it produces an exponentially decaying distribution of chain lengths $N$ [17, 34]:

$$P(N) = \frac{1}{N^*} e^{-N/N^*} \quad , \quad N^* = \sqrt{\phi_0 e^{\Gamma-1}}. \tag{4}$$

Here, $\phi_0$ is the volume fraction of the particles in the suspension, and $\Gamma = \frac{\pi \mu_0 M^2 d^3}{72 k_B T}$ is the ratio of magnetic to thermal energy for the particles (of diameter $d$) in the chain. This simple formula is in good agreement with experimental results over a wide range of conditions [35].

In our case, we are dealing with magnetite particles ($M_S \approx 4 \times 10^5$ A/m [36]) with $d = 21.2$ nm in a suspension where $\phi_0 = 2.8 \times 10^{-5}$. The magnetic field is well above the threshold where $M \approx M_S$, giving us that $\Gamma \approx 20.2$ and $N^* \approx 77$. However, the uncertainty in the average chain length $N^*$ is quite large due to its exponential dependence on $\Gamma$.

It is also possible to estimate the characteristic time for chain assembly within the same theoretical framework. We begin by noting that in our suspensions $\phi_0 \Gamma \approx 6 \times 10^{-4} \ll 0.1$, which means we are in the limit where the assembly rate is set by particle diffusion [17]. The characteristic time for particle-particle binding can then be estimated by the classical Smoluchowsky rate of Brownian particle aggregation [37]:

$$t_B \approx \frac{d^2}{48 \left[\frac{1}{3^{1/2}} - \frac{1}{3^{3/2}}\right] D \phi_0 \Gamma}, \tag{5}$$

For the 3:1 isopropanol/water mixture in which the assembly takes place in this study, the viscosity $\eta = 2.8$ cP [38], and the diffusion constant can be obtained from the Stokes-Einstein equation $D = \frac{k_B T}{3 \pi \eta d} = 74$ µm$^2$s$^{-1}$, yielding that $t_B \approx 5.6$ ms. Note that this value is only valid when there is no long-range interaction between the particles, which means that $t_B$ is a good indicator of chain-building speed only when the solution ionic strength is at least moderately high, so that the long range electrostatic repulsion between particles is screened out. Empirically, the time $t_{\text{assm}}$ to reach a given average chain length follows a power law $N^* \approx C(t_{\text{assm}}/t_B)^{0.6}$ [39]. The constant of proportionality $C$ varies significantly between experiments, but based on previous observations, we can estimate that for chains of the lengths we observe in this work, $t_{\text{assm}} \lesssim 10^4 \, t_B$, or about one minute or less [37, 39, 40].

### 2. Magnetic beads in a rotating magnetic field.

Although SPIONs become ferromagnetic if their size grows beyond about 25 nm, it is possible to create larger superparamagnetic beads by incorporating multiple SPIONs into a larger cluster or bead that is bound together with a magnetically inert matrix consisting, for instance, of poly(acrylic acid) (PAA) [41] or poly(γ-glutamic acid) (PGA) [42]. Such beads can be several µm in diameter. Since they consist of a large collection of SPIONs of random orientations that are fixed relative to each other, they do not have a well-defined easy axis, but respond to an applied magnetic field by acquiring a magnetic moment that will be approximately parallel to the field regardless of its direction. This contrasts to individual SPIONs who need to rotate in order to align their easy axis with the field before their magnetization can be parallel to the field.

Because magnetic beads to a large extent consist of inert binder, their saturation magnetization is typically only 5% - 10% of the SPION $M_S$ [43]. However, since their size is usually about two orders of magnitude larger than our SPIONs, we can see from Eq. (3) that $E_B$ for two µm size beads in contact (scaling as $M_S^2 d^3$) will still be $10^3 - 10^4$ times larger than for SPIONs, and therefore very large compared to $k_B T$. This means that even though the bead assemblies undergo easily visible Brownian motion, thermal fluctuations can be ignored as far as the stability of the magnetic assemblies is concerned. The length of magnetically assembled bead chains will therefore in principle grow without limit unless some other means of capping their length is introduced.

One way to achieve this capping is to perform the assembly in a rotating rather than static magnetic field. Since the chain needs to be aligned with the external field to remain stable, a rotating applied field means that the entire bead chains will rotate at the same frequency $\omega$. This rotation produces a hydrodynamic drag on the particles that acts to break the chain apart, so to find a condition for stability for a chain, it is convenient to introduce the Mason number $M_n$, which is the ratio of the hydrodynamic to the magnetic force on a particle. We define the Mason number as

$$M_n = \frac{16\eta\omega}{\mu_0 M_S^2}. \tag{6}$$

The speed with which a particle in a chain moves through the fluid equals $v_p = \omega l$, where $l$ is the distance to the center of the chain. Consequently, drag forces are larger the farther from the center particles are located. Moreover, every particle in the chain experiences at least some drag, and this propagates through the chain, so that the central bond in the chain experiences a fragmenting force from all the particles in the chain. This force then grows as the square of the chain length $N$, and when it becomes larger than the magnetic bonding force, the chain will break in the middle, which sets a cap for the maximum obtainable chain length. Gao et al. [20] modelled this behavior, showing that stability of a chain with $N$ particles can be parametrized by

$$R_T = \frac{M_n N^3}{(N-1)\left(\ln\frac{N}{2} + \frac{2.4}{N}\right)} \tag{7}$$

If a chain is long enough that $R_T > 1$, it is unstable and breaks apart, while all shorter chains are stable. There is good agreement between this equation and experiments performed on magnetic beads at the µm scale [20, 44, 45], as well as with numerical simulations [46].

The derivation of $R_T$ neglects any effect of temperature and any forces other than magnetic attraction and hydrodynamic drag. We have already noted that this is reasonable for larger beads, but less so for the nanoscale SPIONs we study here. If those effects were included in the model, chains would be more prone to fragmentation, and we could no longer use the value of a parameter such as $R_T$ to issue a binary stable/unstable prediction for a given chain, as chains of any length are in principle subject to fracture due to thermal fluctuations. Consequently, we would expect the maximum chain length for smaller particles to be less than predicted by Eq. (7).

### B. Experiments on SPIONs in a rotating field

Our experiments were performed on commercially obtained SPIONs that were diluted in an aqueous solution to a concentration of approximately 0.14 mg/ml. The particle diameter was nominally 20 nm, but a TEM study of the particles used (see Fig. 2), gives us a mean diameter of 21.2 nm, which was used in all calculations.

To ensure that the magnetic attraction between the particles was sufficient to overcome the electrostatic repulsion between their citrate-coated surfaces, 0.5 mM of NaCl was also added to the SPION suspension. A vial of this suspension was placed between two parallel rare earth block magnets and made to rotate at rotational speeds between 600 rpm (10 Hz) and 3600 rpm (60 Hz) while the magnets were kept stationary, as shown schematically in Fig. 1(a). The magnetic field in this space was measured to be 0.44 T, well beyond what is required to achieve full saturation of the magnetization in the SPIONs, causing them to rapidly assemble into chains. At the same time, since the permanent magnets are about three times as wide as their separation, the magnetic field at the center of the field space is sufficiently uniform that any drift of the chains due to gradient forces can be neglected.

Nanoparticles have the disadvantage that their small size preclude easy study with optical microscopy. Therefore, after the chains had been allowed to evolve for up to 2 hours, a solution of TEOS (tetraethyl orthosilicate) was added to the suspension, causing a silica shell to form around the SPION chains. This reaction was allowed to proceed for 3 hours while maintaining the rotation of the vial. With the silica shell in place, the chains remained intact when the suspension was removed from the magnetic field, so they could be imaged with transmission electron microscopy (TEM). Examples of TEM micrographs from samples fabricated with six different values of $\omega$ are shown in Fig. 3(a). We can clearly see that the chains that form become progressively shorter as $\omega$ is increased, but also that each sample contains a wide range of chain lengths. In Fig. 3(b) we show closeups of short and long chains from the 600 rpm sample. We see from this that while the chains are largely linear, they exhibit significant disorder at small scales.

To produce quantitative data on the samples, we measured the lengths of all fully visible chains in several TEM images from each sample, counting between 70 and 200 chains in each case. All the micrographs contained very large numbers of few-particle chains and single particles, so to make this a manageable task, we confined ourselves to counting only chains with 6 or more particles.

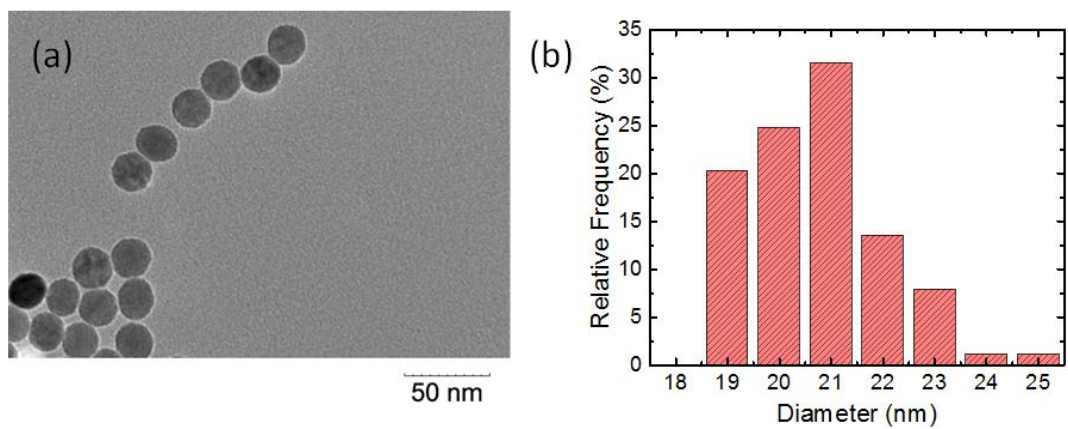

*Figure 2: (a) TEM image of as-received SPIONs. (b) Size distribution obtained through TEM imaging, yielding an average diameter of 21.2 nm.*

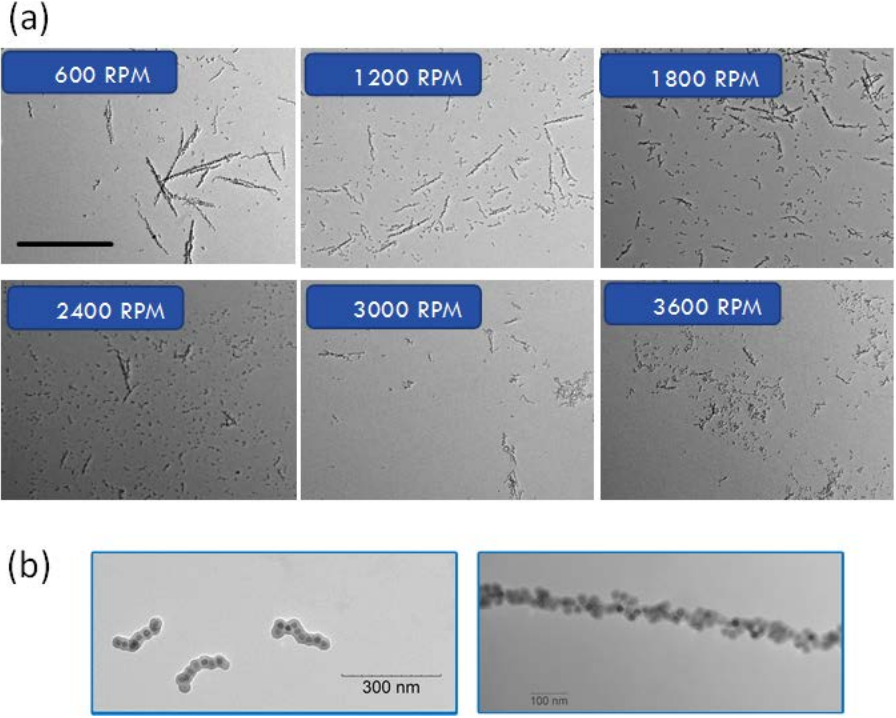

*Figure 3: (a) A series of TEM images of SPION chains with different rotational angular velocities ω (600, 1200, 1800, 2400, 3000, 3600 rpm). Salt concentration equals to 0.5mM in all cases. (b) Representative images of short and long chains selected from the 600 rpm sample.*

Physical chain lengths were converted to normalized length $L_n$ by dividing the lengths by the average diameter of the SPIONs. Fig. 4(a) plots the measured distributions obtained for the highest and lowest values of $\omega$ (600 rpm and 3600 rpm, respectively shown as grey and red bars). The distributions appear approximately exponential, so they were fit to the distribution in Eq. (4), replacing the average chain length $N^*$ with a free parameter $M^*$ that was determined by the fit. In addition, we took the 98th percentile $N_{98\%}$ of each distribution as a stand-in for the maximum obtainable chain length $N_{\max}$. Both $M^*$ and $N_{98\%}$ are plotted against $\omega$ in Fig. 4(b) along with curves of chain lengths corresponding to constant values of $R_T$ from Eq. (7).

The exponential chain length distribution indicates that thermal fluctuations play an important role in the process of chain fragmentation. The calculated $N^*$ for our experiments is about 77, which should set an upper bound on the average chain lengths we observe, and indeed we find that $M^* \ll N^*$ in all cases.

Next, we turn our attention to the very longest chains we observe in the experiments. If we ignore temperature and interactions such as electrostatic repulsion between particles, they should form at $R_T \approx 1$, but this is very clearly not the case. In fact, the 98th percentile of our distribution is a rather good fit to $R_T = 0.025$, which corresponds to chains that are only one 6th the length of chains with $R_T = 1$. Again, this demonstrates the important role of thermal fluctuations in causing fragmentation of the SPION chains we study here. Clearly, a complete theory for the assembly process must take temperature into account for magnetic particles of this size. It is interesting to note that in spite of the importance of temperature in our case, the $R_T = 0.025$ curve is a good predictor of maximum chain length over the range of $\omega$ values we are able to study here, indicating that the $1/\sqrt{\omega}$ dependence of chain length on $\omega$ implied by Eqs. (6) and (7) still holds when thermal effects become important.

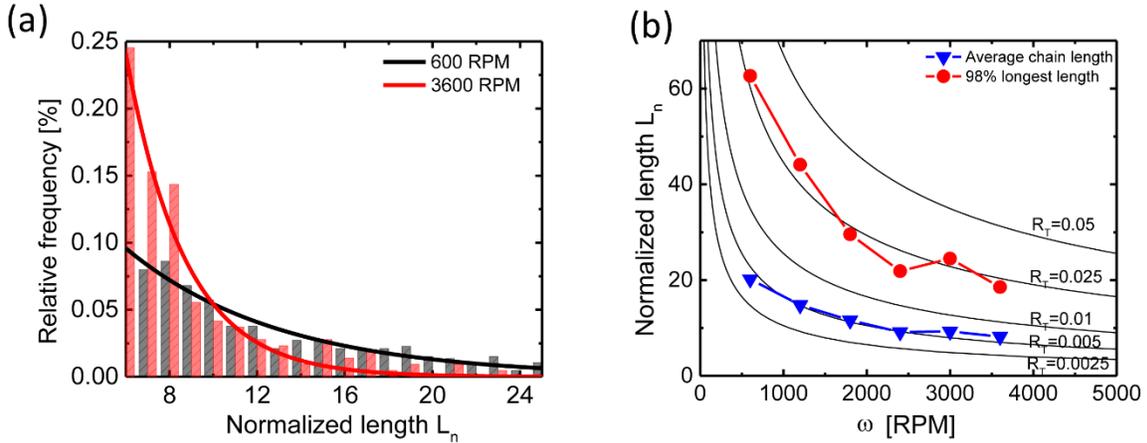

*Figure 4: (a) Length distribution of SPION chains assembled under two different rotational speeds (600 RPM and 3600 RPM), both fit to exponential distributions. (b) Plot of average length ($M^*$, blue triangles) and 98% longest length ($N_{98\%}$, red disks) for chains fabricated at different values of ω. The black lines indicate constant values of the parameter $R_T$, as defined in the text.*

### C. Disorder in SPION chains.

Disorder is one of the main impediments to the efficient and effective self-assembly of nanoparticles into useful structures, and developing a better understanding of it is therefore a high priority [26, 47]. There are many reports of a high degree of disorder in SPION chains [28, 48, 49], which stands in contrast to the state of affairs for chains of microbeads, which tend to exhibit very low disorder [20, 44, 45]. This difference in behavior has not been fully explained, but we can identify at least two factors that are important. The first of these is again the fact that thermal fluctuations are much more important for the small SPIONs than for the larger beads. This means that Brownian motion plays a crucial role in SPION assembly, making it likely that two closely separated particles are pushed together into a configuration that does not correspond to a minimum in free energy, where friction between the particles can easily trap them. In the larger beads, by contrast, the assembly will be characterized by slow drift along magnetic field lines, making it likely that the aggregating particles will only come into contact when they are relatively close to their optimum alignment.

The second factor that may be important to the difference in disorder between the two systems is the lack of an easy axis in the larger magnetic beads. This property means that if a bead attaches to a chain away from the chain's most distal point, the bead can correct its misalignment by simply rolling towards its optimal position. In SPIONs, rolling cannot occur because the easy axis is both fixed in the particle and closely aligned with the external magnetic field during the assembly. The SPION therefore maintains its orientation at all times, and must slide rather than roll if a misaligned particle is to attain the position of minimum energy. The chain of SPIONs will therefore in general experience a much high friction barrier against the healing of disorder than a chain of beads. Consistent with this picture, Monte Carlo simulations indicate that disorder in SPION chains can be significantly reduced if this friction is lowered [50].

Disorder is clearly evident in the TEM images in Fig. 3, and as with the case of the chain length distribution, quantifying this disorder will require us to characterize the disorder in a large number of chains. To do this, we must first settle on a useful and robust measure of disorder in the chains that is also not too cumbersome to obtain from the TEM images. A good number of disorder measures exist in the literature [51, 52], but we have elected to use the normalized length $L_n$ because it is straightforward to calculate, and provides a sensitive measure of disorder in chains with low to moderate disorder [50]. $L_n$ for an aggregate of particles is given by

$$L_n = \frac{D_{\max}}{d(N-1)}, \qquad (8)$$

where $D_{\max} = \max_{ij}\sqrt{(r_i - r_j)^2}$ is the maximum distance between any two particles in the aggregate. $i, j$ here run over the positions $r$ all the $N$ particles in the chain, and $d$ is the hydrodynamic radius of a single SPION. $L_n = 1$ in a perfectly colinear chain where there is no overlap between particles, and any number smaller than this indicates some degree of disorder. Finding $L_n$ for our chains then only requires us to measure the chain length (which is $\approx D_{\max}$) and to count the number of particles in the chain.

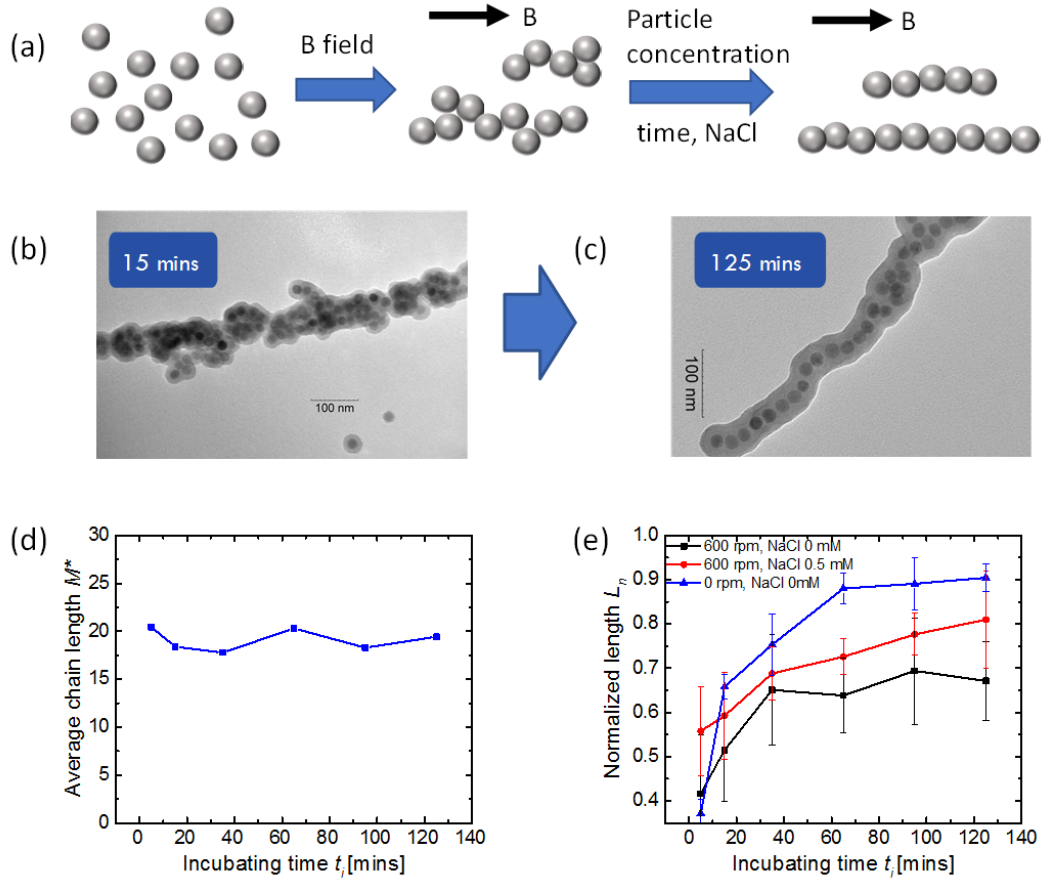

Figure 5: (a) Cartoon of the disorder dissipation process. (b) TEM of a representative chain after 15 min incubation. (c) Same, but for 125 min incubation. (d) Average chain lengths vs incubation time. (e) Normalized length $L_n$ (quantifies chain order) vs incubation time for three different experimental conditions.

1. *Effect of incubation time on disorder*

We hypothesized that if a chain is given enough time before it is made permanent with silica encapsulation, some of the disorder that is present on initial assembly may dissipate as particles find ways, due for instance to Brownian impulses, to reach a configuration closer to the energy optimum. This process is indicated schematically in Fig. 5(a). To test it, we performed a series of experiments where the time between insertion into the magnetic field and introduction of TEOS was varied between 5 minutes and 125 minutes. The encapsulation and TEM imaging were implemented as previously, resulting in TEM images such as those in Fig. 5(b) and 5(c). The difference between the chains that have been incubated for 15 minutes and 125 minutes is striking. After 15 minutes, the chain is highly disordered, to the point where it is approximately three particles wide, sporting a variety of protrusions, while after 125 minutes have passed, the chain is only one particle wide, containing stretches of nearly ideal colinear order. Varying the incubation time has no measurable effect on the chain length distribution, as indicated by the data in Fig. 5(d), which means that if disorder has any effect on the cohesion of the chains, that effect is too small for us to observe.

In Fig. 5(e) we have plotted the average values of $L_n$ from the fabricated samples, where each data point represents an average over at least 25 chains. As the TEM images suggest, we see a distinct increase in order as the incubation time is lengthened, with a particularly steep improvement in the first 40 minutes. This is a very long time compared to $t_B$, which as we have already seen is about 5.6 ms, and also long compared to our estimate for the longest time to assemble the full-length chains (less than one minute). In other words, the portion of the disorder that can heal does so over times that are on the order of $10^5$ times longer than the characteristic time scale for SPION particle assembly. It is also worth noting that changes in incubation time of as little as 10 min have clearly noticeable effects on chain disorder. This is in spite of the fact that we are allotting three full hours after the incubation for the silica layer to grow to completion. This behavior makes sense if we assume that the growth of the silica layer stabilizes the chains of a time scale much shorter than three hours. In that case, the evolution of chain configurations comes to a rapid stop after TEOS injection, providing a distinct end to the incubation period. This notion is supported by the fact that in most cases, the separation between SPIONs in a chain is small compared to the silica layer thickness (but see Fig. 5(b)), indicating that once the silica has grown a certain amount, further magnetic assembly becomes unlikely.

Fig. 5(e) contains three traces corresponding to different experimental conditions, but that all show the same trend of improved order over time. The trace that corresponds to chain assembly at $\omega = 0$ rpm (blue line with triangular markers) shows the greatest increase in order, saturating at $L_n$ of 0.9, clearly higher than the other traces that were assembled at $\omega = 600$ rpm. In other words, the rotating magnetic fields partially impedes the healing of disorder in the chains. This may be because rotating chains have a chance of undergoing high-impact collisions, while chain-chain interactions are going to be much gentler in the non-rotating case. If the collisions have a chance of decreasing order in the chains, that could explain the observation. We also see that adding salt to the suspension during incubation improves order compared to the case of no salt added. This is an important clue to the mechanism behind the disorder dissipation and will be treated next.

### 2. Effect of salt concentration on disorder

We are not yet at the stage where we can confidently identify the mechanism by which the disorder heals, but the data contains clues which may point toward an answer. For instance, Fig. 5(e) show results of experiments performed without salt in the suspension (black trace with square markers) and with some salt ($c_{NaCl} = 0.5$ mM, red trace with circular markers). It is clear that the salt reduces the disorder at all tested values of $t_i$, and also that it continues to improve for a longer period of time than in the no-salt case, where $L_n$ appears to saturate after about 40 minutes.

To further investigate this highly unexpected result, we performed a series of measurements where we kept $t_i$ constant at 15 min and varied the salt concentration from 0 mM to 1.5 mM. TEM images of chains from the end points in this series are shown in Fig. 6(a) and 6(b), and values of $M^*$ and $L_n$ obtained from each ensemble of chains in each sample are plotted in Fig. 6(c). For $M^*$, we see a slight increase in chain lengths with increasing salt concentration, which is reasonable given that the salt will partially screen out the electrostatic repulsion between particles, leading to a stronger bond between magnetically assembled particles.

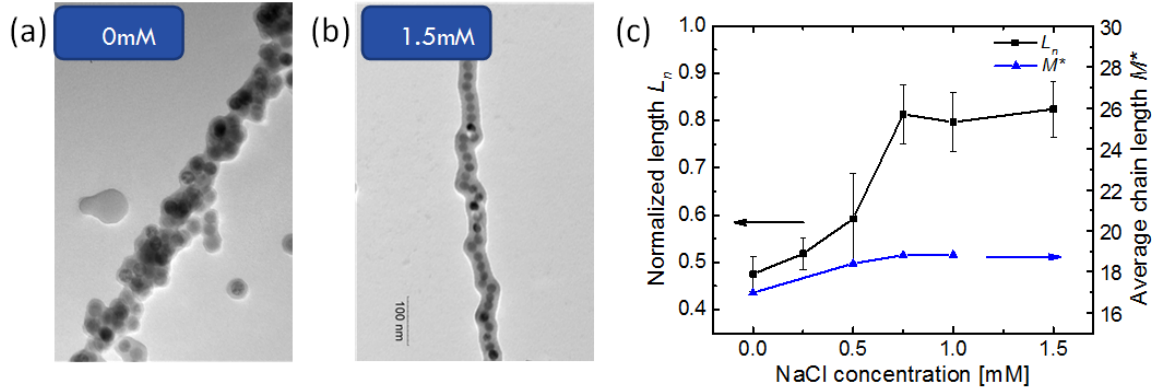

Figure 6: (a) Representative TEM micrograph of chain assembled with $t_i = 15$ min, without salt. (b) Same, but with 1.5 mM NaCl in solution. (c) Normalized length $L_n$ and average chain length $M^*$ vs salt concetration, with $t_i = 15$ min. Chain length was not calculated for the largest salt concentration due to a high level of chain aggregation on the TEM grid.

The trend in Fig. 6(c) is quite similar to what we saw in Fig. 5(e); the larger the salt concentration, the greater the order of the chains, with a saturation in $L_n$ around ~0.8 at the higher salt concentrations. In the TEM images, Fig 6(a) corresponds to the same conditions as in Fig. 5(b)—no added salt and a short (15 min) incubation, and the chains therefore exhibit the same amount of disorder. Chains incubated for the same short time at 1.5 mM salt concentration (Fig. 6(b)) exhibit lower disorder, resembling chains incubated for 125 minutes at 0.5 mM salt concentration (Fig. 5(c)).

The improvement in order appears to saturate near a NaCl concentration of 0.7 mM. In the alcohol/water mixture of the suspension, this corresponds to a Debye length $\kappa^{-1}$ of approximately 7.7 nm, which is slightly smaller than the radius of the SPIONs. This scale may offer a clue to the mechanism behind the salt-mediated improvement in order. To see this, we model the particle-particle interaction with a modified DLVO [53] potential that also includes the magnetic interaction:

$$V_{DLVO}(r) = 2\pi a \epsilon \psi^2 \ln\bigl(1 + e^{-\kappa c_{NaCl}(r-2a)}\bigr) - \frac{A}{6}\left(\frac{2a^2}{r^2 - 4a^2} + \frac{2a^2}{r^2} + \ln\left(1 - \frac{4a^2}{r^2}\right)\right)$$
$$- \frac{\mu_0 M_S^2 V^2}{4\pi r^3}(3\cos^2\theta - 1) \qquad (9)$$

The first and second terms in this equation account for electrostatic repulsion and van der Waals attraction between particles, while the third term, which can be either repulsive or attractive, is the magnetic potential $E_B$ from Eq. (3). $\psi$ is here the SPION electrostatic surface potential, $a = 10.6$ nm is the average SPION radius, and $A$ is the nonretarded Hamaker constant for magnetite. Since the experiments are performed in a 3:1 mixture of isopropanol and water, we use $\epsilon = 35\epsilon_0$ for the dielectric constant of the solvent [38]. This lower $\epsilon$ affects the strength of the nonmagnetic interactions in Eq. (9). For the surface potential, we measure a zeta potential of = -35 mV for our SPIONs when dispersed in low ionic strength water. To accommodate the presence of alcohol, we use the Grahame equation [54], to estimate that $\psi_0 = -66$ mV under the assumption that the particle surface charge is the same in the alcohol/water mixture as in pure water. The Debye length $\kappa^{-1}$ is proportional to $\epsilon^{-\frac{1}{2}}$, and is therefore about one third lower in the alcoholic mixture than in water. We also expect that the Hamaker constant would be somewhat lower in the alcoholic mixture, but since $V_{DLVO}$ is largely unaffected by $A$ beyond particle separations of a few nm, this is of less importance to our results, and we will use the value in water $A = 3.3 \times 10^{-20}$ J in our calculations [55].

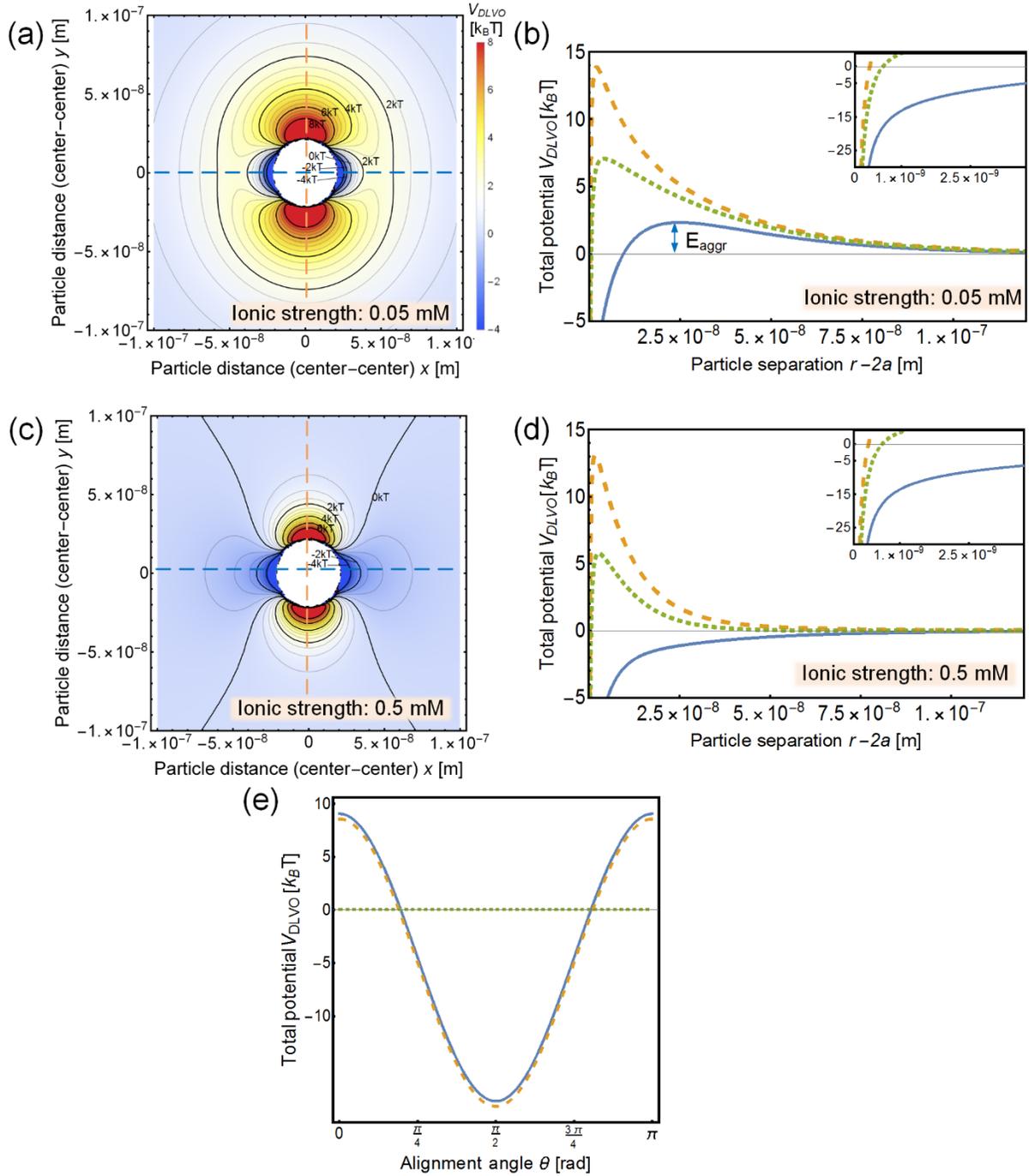

*Figure 7: DLVO model calculations of total interaction potential $V_{DLVO}$ for SPIONs in a strong magnetic field. (a) Ionic strength $I = 0.05$ mM. The x-axis corresponds to end-to-end alignment, and the y-axis to side-to-side-alignment. (b) Traces of $V_{DVLO}$ along the x-axis (blue solid line) and y-axis (orange dashed line) in (a), as well as $V_{DLVO}$ in the absence of an external magnetic field (green short-dashed line). (c) and (d): As (a) and (b), but with $I = 0.5$ mM. (e) $V_{DLVO}$ plotted for a particle separation of 0.6 nm for ionic strengths of $I = 0.01$ mM (blue solid line) and 1.0 mM (orange dashed line). The green dotted line indicates the interaction potential at 0.6 nm separation without a magnetic field present.*

In Fig. 7, we plot $V_{DLVO}$ as calculated for ionic strength $I = 0.05$ mM (Fig. 7(a) and 7(b)), and for $I = 0.5$ mM (Fig. 7(c) and 7(d)). In both cases, we see that there is a deep minimum when two SPIONS are in contact and their magnetic moments are aligned end-to-end ($x = 21.2$ nm, $y = 0$ nm), as expected. However, for the lower ionic strength, there is an aggregation barrier of height $E_{\text{aggr}} \approx 2.3\ k_B T$ that needs to be overcome before magnetic binding between particles can take place. This barrier disappears completely at $I = 0.22$ mM, and for higher salt concentrations, the region of attractive interactions increases in volume with increasing ionic strength until $I \sim 0.5$ mM. As a result, it is easier for particles to bind to each other at higher solution ionic strength, and the growth of chains will therefore occur at a faster rate. This is the limit in which the Smoluchowski particle binding time $t_B$ given by Eq. (5) is likely a good indicator of chain formation speed. For ionic strengths much lower than 0.05 mM, $E_{\text{aggr}}$ rises substantially. For $I = 0.01$ mM, we obtain $E_{\text{aggr}} \approx 4.5\ k_B T$, which should prevent or at least significantly slow chain formation. However, no particular efforts were undertaken to remove all salt from the solution. It is therefore reasonable that all samples tested had $I$ not much less than 0.05 mM.

At the same time, even the highest ionic strengths are not high enough that they substantially affects the interaction between particles that are in contact or very nearly in contact. This is illustrated in Fig. 7(e), were we plot $V_{DLVO}$ vs alignment angle $\theta$ for particles separated by 0.6 nm, and at ionic strengths of 0.1 mM and 1.0 mM. It is clear that the salt produces only a minor perturbation of the energy landscape once the particles have bound to each other. This point is also supported by the fact that the average chain length rises only slight as $c_{NaCl}$ is increased from 0 to 1.0 mM (blue trace in Fig. 6(b)).

In other words, a salt concentration on the order of 0.22 mM is large enough that it suppresses the long-range electrostatic repulsions between particles, allowing them to more easily come within the range the magnetic attraction, but not so large that there is an appreciable modification of interparticle interaction when the particles are in contact. The main effect of the salt is therefore to speed up chain growth without substantially altering the nature of the kinetic traps that are the cause of chain disorder.

This observation points to the possibility that the process of chain growth is important to the healing of chain disorder. This is a reasonable conclusion considering that a sizeable external energy must be provided to dislodge a particle that is kinetically trapped in a deep local minimum. If the trap is deeper than a few $k_B T$, thermal fluctuations are unlikely to provide this energy, even at long time scales. However, as a chain grows through the addition of more particles, the forces within the chain will shift, changing the energy landscape, potentially releasing some kinetic traps and thereby giving the chain an opportunity to move closer to the global energy minimum. In addition, the chain growth will also regularly result in the scission of chains, which may provide jolts that similarly can dislodge chains from kinetic traps.

We should also mention that DLVO theory predicts that aggregation in a magnetic field occurs at the primary minimum corresponding to direct contact between the particles. For larger particles, the theory predicts that magnetic binding should predominantly occur at the secondary minimum where particles are not in direct contact, but instead separated by a small distance [56], but for particles of the smaller scale we use here, the secondary minimum is very shallow and exists only over a narrow range of salt concentrations. Since the primary minimum is always attractive regardless of any magnetic field, one might expect that magnetic binding would be irreversible in SPIONs, which is contrary to observations. Most likely, there is a thin steric layer surrounding the SPIONs preventing irreversible aggregation from taking place. With the chosen interaction strengths in our model, this layer can be as thin as 0.3 nm to prevent irreversible aggregation from occurring (as illustrated by the green line in Fig 7(e)). Since the value of $A$ may be lower than what we have used in our calculations, the minimum thickness of the steric barrier may in fact be even lower than this.

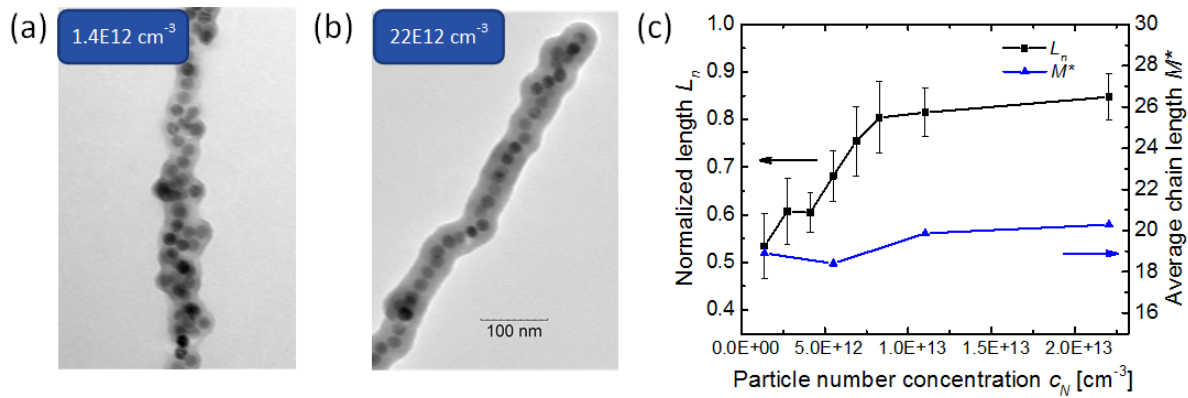

*Figure 8: (a) Representative TEM micrograph of chain assembled with $c_N = 1.4 \times 10^{12} cm^{-3}$. (b) Same, but with $c_N = 22 \times 10^{12} cm^{-3}$. (c) Normalized length $L_n$ and average chain length $M^*$ vs particle concentration. $t_i = 15$ min in all cases.*

### 3. Effect of particle concentration on disorder

If our hypothesis is correct, then any mechanism that increases the "churn" in the chain assembly will have the effect of speeding up the dissipation of chain disorder. With this in mind, we next investigated the effect of an increased particle concentration on $L_n$. The particle number concentration $c_N$ in all experiments described up to this point was $5.5 \times 10^{12}$ cm$^{-3}$. This was now varied between $1.4 \times 10^{12}$ cm$^{-3}$ and $22 \times 10^{12}$ cm$^{-3}$, while salt concentration was maintained at 0.5 mM and $t_i$ was fixed to 15 min. Fig. 8(a) and (b) display TEM images of chains obtained at the lowest and highest concentration, Fig. 8(c) plots $M^*$ and $L_n$ as a function of $c_N$.

The TEMs in Fig 8(a,b) have a very similar appearance to their counterparts in Figs. 5 and 6, and we again see an increase in $L_n$ from a minimum in the 0.4-0.5 range to a maximum around 0.8-0.9, with a saturation occurring near the middle of the range, in this case around $8 \times 10^{12}$ cm$^{-3}$. In other words, we see precisely the behavior predicted by our hypothesis. On the other hand, $M^*$ increases very slowly or not at all with $c_M$, in apparent contradiction to the prediction of Eq. (4).

### 4. Effect of particle surface on disorder dissipation.

As a final component of this study, we investigated the effect changing the surface properties might have on chain disorder. Since we are already using silica coating for chain encapsulation, a convenient way to modify the SPION surface is to encapsulate the particles in silica prior to exposure to the magnetic field. This was done in the same manner as the chain encapsulation. The result is a 4 nm thick silica coating on each particle in the starting suspension, as shown in Fig. 9(a). The coating also results in an increase in zeta potential in water to -55 mV. We performed a series of experiments on these modified particles, measuring chain length and disorder as a function of incubation time and salt concentration, as shown in Fig. 9(c) and (d). The first clear difference from the earlier results is that $M^*$ is smaller by more than a factor of two. This is expected, since the approximately 50% increase in particle radius without any change in magnetic moment gives us a magnetic binding energy $E_B$ that is only about 30% of the value for the uncoated SPIONs. In fact, from our initial calculations of $\Gamma$ and $N^*$ we might have expected an even greater reduction in chain length, but given the large uncertainty of that calculation we will not draw any inferences from this discrepancy.

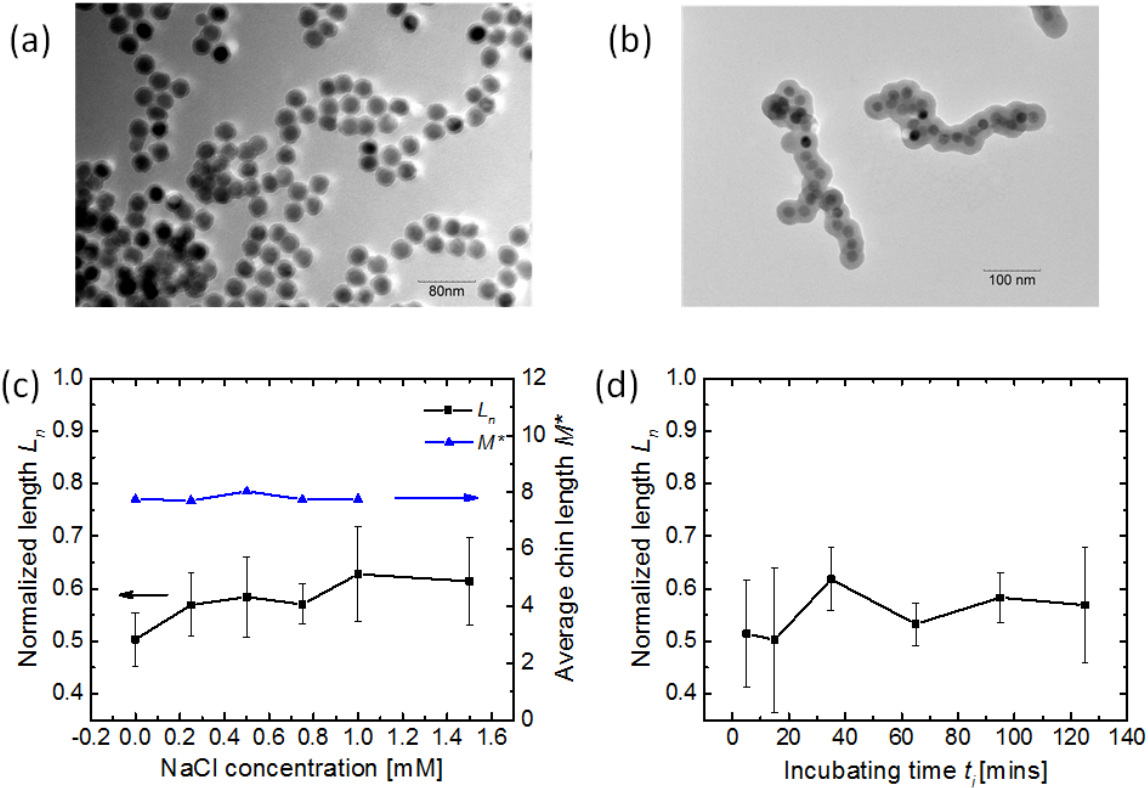

*Figure 9: (a) SPIONs coated with 4 nm silica. (b) Representative TEM micrograph of chain assembled from silica coated particles at 600 rpm, with $t_i = 15$ min and 0.5 mM NaCl in solution. (d) Normalized length $L_n$ and average length $M^*$ as a function of salt concentration. (e) $L_n$ of silican-coated SPIONs as a function of incubation time.*

As for the disorder in these shorter chains, it starts out high and is unaffected by incubation time (Fig 9(d)), and barely affected by salt concentration (Fig. 9(c)). This is in spite of the lower magnetic interaction energy, which should lead to a corresponding decrease in the depth of local traps. The high disorder in these chains is therefore likely due to a higher friction between silica-coated particles compared to the as-prepared particles, preventing the dissipation of the disorder, or at least significantly slowing it down.

## III. CONCLUSION

We have investigated chaining of SPIONs in a rotating magnetic field, a regime that has up to now been largely overlooked, although the two limiting cases of SPION assembly in a static field and the assembly of micron-sized magnetic beads in a rotating field are both well studied. The resulting ensemble of SPION nanochains retains features of the better understood limiting cases, in particular an exponential distribution of chain lengths seen in the static field limit, as well as a maximum chain length that scales as $1/\sqrt{\omega}$, just as is the case in the large particle limit. However, the chains are shorter than predicted by theories developed for either limit. This reduction occurs because while hydrodynamic drag is absent in the static field case, and thermodynamic fluctuations can be ignored for large particles, both effects must be taken into account in the regime we have studied here, where both operate to reduce the length of the nanochains. To develop a theory that incorporates both effects one would need to move beyond the realm of near-equilibrium thermodynamics. This is likely to be challenging, but if mastery of self-assembly at the nanoscale is to be accomplished, it is crucial that such models and theories be developed. The system we have investigated is both relatively simple and accessible to experiments, and may therefore be a suitable target for such work.

The fabricated nanochains display varying degrees of disorder, but this disorder interestingly does not seem to appreciably affect the distribution of chain lengths or the relationship between field rotation frequency $\omega$ and maximum chain length. This suggests that the detailed configurations of the SPION chains have relatively little impact on the steady-state balance between chain growth and fragmentation, which may offer a clue to how theoretical models of the system might be simplified without losing predictive power.

As for the disorder itself, we find that if the chain suspension is left in the rotating field after its steady state length distribution has been achieved, disorder gradually dissipates over a timescale of tens of minutes, about two orders of magnitude slower than the chain assembly time. The improvement in order can be sped up by increasing either ionic strength or particle concentration, both of which increases the speed of the chain assembly process. This result strongly suggests that the growth and scission of chains that continually takes place during the experiment is responsible for the improved order. Further studies are needed to pin down the mechanism more precisely, but it is highly plausible that the gradual amelioration of disorder over time is in large part due to shifts in the energy landscape within the chains brought about by the gradual lengthening of the chains as they merge with particles and other chains. This all occurs in steady state, where chains break up at a rate that precisely balances the rate of chain growth, which is why the time scale for disorder healing is so much slower than the time required to reach a stable chain length distribution.

Our disorder studies clearly show that although chains of SPIONs are highly prone to disorder through kinetic trapping, there are ways to significantly improve the order of the chains without necessarily modifying the surface of the particles to minimize friction. This result points to what could become a general method for relieving kinetic trapping in nanoassembled systems. The standard approach to this problem is to try to avoid it altogether by ensuring that there are no trap states in the configuration space of a nanostructure that are deeper than a few $k_B T$, so that thermal fluctuations eventually guide the system to the desired state. For SPION chains, this could be accomplished by lowering the friction between particles sufficiently. Our results show that disorder can sometimes be reduced by operating away from thermal equilibrium, where non-thermal energy can be supplied toward the release of kinetic traps in self-assembled structures, allowing them to evolve toward the global free energy minimum. Since kinetic trapping is a very important issue in self-assembly, it would be very interesting to explore whether this idea could be generalized to other self-assembled systems. As a first step, further studies of the SPION chains are called for, so that we can gain a more detailed understanding of the physics of this relatively simple model system.

We need to point out that this effect is no panacea against disorder. We have not observed $L_n$ greater than about 0.90 in any of the samples we fabricated, which still leaves a substantial amount of disorder in the chains, as can be seen through Figs. 5, 6 and 8. Compared to the near-perfect collinearity seen in chains of magnetic microbeads, this leaves a lot to be desired. Still, the disorder dissipation we do observe corresponds to a redistribution of particles that cuts the number of SPIONs required to form a chain of a given length nearly in two, and it is possible that with further refinement of the techniques, we can get even closer to perfectly ordered chains.

## IV. Materials and methods

### A. Preparation of citric acid-coated SPIONs

40 µl of a suspension of oleic acid-coated magnetite nanoparticles in chloroform (20 nm nominal size, 1.72 mg/ml, Ocean Nanotech) were dispersed in 2 ml *N, N'*-dimethylformamide (99%, Fluka Chemicals), to which 0.1 g of citric acid (99.5%, Acros Organics) was added [57]. The solution was then vigorously stirred at 95 °C for 6 h. The particles were separated magnetically with a small permanent magnet held to the beaker, and washed in nanopure water. To fully remove any free citric acid, this step was repeated three times, after which the particles were redispersed into 2ml nanopure water, resulting in a suspension with a nominal nanoparticle concentration of 700 µg/ml. Transmission electron microscopy was used to determine that the average diameter of the particles was 21.2 nm.

### B. Fabrication of the SPION nanochains

To coat the chains, a modified Stöber method was employed [18, 25]. Tetraethylorthosilicate (TEOS, 99%, Sigma-Aldrich), ammonium hydroxide (NH$_4$OH, 30%, Spectrum Chemical), and Isopropyl alcohol (IPA, 99.7%, Spectrum Chemical) were used as reagents. First, 320 µl IPA, 80 µl NH$_4$OH (3% v/v in water) was added to 100 µl of an aqueous suspension of citric acid-terminated magnetite nanoparticles, prepared as just indicated. An aliquot of 50 mM NaCl in water solution was also added to adjust the salt concentration to between 0 mM to 1.5 mM. The mixture was transferred to a 1.0 ml vial, and placed in a PEEK plastic chuck mounted at the center of a magnetic field space consisting of two parallel 2"×2"×0.25" N52 Neodymium block magnets mounted at a separation of 0.75" within a yoke made from gray cast iron. The space provides a 0.44 T uniform magnetic field at the location of the particle suspension.

The sample was allowed to incubate at rest for 5 minutes in the magnetic field, at which point a Pittman 1312S103-SP Brushless DC Motor, connected with direct drive to the chuck, was spun to an angular velocity ranging from 600 rpm to 3600 rpm and held at that speed for between 0 and 120 mins. We refer to his time, plus the 5 minutes initial incubation, as the total incubation time $t_i$. 40 µl of a solution of TEOS (2.25 mM in IPA) was injected into the sample vial, producing a 90 µM TEOS concentration. The reaction was allowed to proceed to completion under rotation for 3 hours, resulting in a silica shell, approximately 10 nm thick, forming on all chains in the suspension. The suspension was cleaned by centrifugation at 200 rcf for 20 minutes, after which the supernatant was removed and nanopure water was added to restore the initial volume. The suspension was immediately drop cast onto a TEM grid and allowed to dry completely before imaging.

## ACKNOWLEDGEMENTS

The authors gratefully acknowledge prof. Richey Davis for helpful discussions.